\def\year{2022}\relax
\documentclass[letterpaper]{article} % DO NOT CHANGE THIS
\usepackage{aaai22}  % DO NOT CHANGE THIS
\usepackage{times}  % DO NOT CHANGE THIS
\usepackage{helvet}  % DO NOT CHANGE THIS
\usepackage{courier}  % DO NOT CHANGE THIS
\usepackage[hyphens]{url}  % DO NOT CHANGE THIS
\usepackage{graphicx} % DO NOT CHANGE THIS
\usepackage{booktabs}
\usepackage{amssymb} 
\usepackage{natbib}  % DO NOT CHANGE THIS AND DO NOT ADD ANY OPTIONS TO IT
\usepackage{caption} % DO NOT CHANGE THIS AND DO NOT ADD ANY OPTIONS TO IT
\DeclareCaptionStyle{ruled}{labelfont=normalfont,labelsep=colon,strut=off} % DO NOT CHANGE THIS
\usepackage{algorithm}
\usepackage{algorithmic}

\usepackage{newfloat}
\usepackage{listings}
\floatstyle{ruled}
\newfloat{listing}{tb}{lst}{}
\floatname{listing}{Listing}
\title{Representation Learning for Wearable-Based Applications\\ in the Case of Missing Data}
\author {
    Janosch Jungo\textsuperscript{\rm 1*},
    Yutong Xiang\textsuperscript{\rm 2*},
    Shkurta Gashi\textsuperscript{\rm 2, 3},
    Christian Holz\textsuperscript{\rm 2, 3}
}
\affiliations {
    \textsuperscript{\rm 1} Department of Information Technology and Electrical Engineering, ETH Zürich, Switzerland\\
    \textsuperscript{\rm 2} Department of Computer Science, ETH Zürich, Switzerland \\
    \textsuperscript{\rm 3} ETH AI Center, ETH Zürich, Switzerland\\
    \textsuperscript{\rm *} These authors contributed equally. \\
    jjungo@student.ethz.ch, xiangy@student.ethz.ch, shkurta.gashi@ai.ethz.ch, christian.holz@inf.ethz.ch
}
\usepackage{bibentry}
\begin{document}

\maketitle

\begin{abstract}
Wearable devices continuously collect sensor data and use it to infer an individual's behavior, such as sleep, physical activity, and emotions. 
% Such information can be used by computing systems to help people improve their health and overall well-being. 
Despite the significant interest and advancements in this field, modeling multimodal sensor data in real-world environments is still challenging due to low data quality and limited data annotations. In this work, we investigate representation learning for imputing missing wearable data and compare it with state-of-the-art statistical approaches. We investigate the performance of the transformer model on 10 physiological and behavioral signals with different masking ratios. Our results show that transformers outperform baselines for missing data imputation of signals that change more frequently, but not for monotonic signals. We further investigate the impact of imputation strategies and masking rations on downstream classification tasks.
% and discover insightful findings related to the sensor and setting of the data collection.
% The proposed method is suitable for the majority of sensor types and for medium or long missing sequence lengths. 
Our study provides insights for the design and development of masking-based self-supervised learning tasks and advocates the adoption of hybrid-based imputation strategies to address the challenge of missing data in wearable devices.
\end{abstract}

\section{Introduction}

Personal computing devices -- such as smartwatches and smart rings -- and their wide range of sensors -- such as heart rate and accelerometer -- provide an unprecedented opportunity to understand and infer people's behavior, e.g., physical activity \cite{Radu2018_MultimodalDL, Ordonez2016_DeepConvLSTMHAR, Guan2017_EnsemblesDeepLSTMHAR}, emotions \cite{Picard2000_AffectiveComputing, Sano2013_StressDetection}, sleep \cite{Zhai2021_UbiSleepNet, Zhai2020_MakingSenseOfSleep}, and overall health \cite{Adao2021_FatigueMonitoring, Tong2019_TrackingFatigueMS}. However, variability of the data across subjects, devices, and locations along with the scarcity of labeled data and low data quality pose significant challenges \cite{Plotz2021ApplyingMLforSensorData}.

To address these limitations, several researchers leveraged self-supervised learning (SSL) to learn directly from unlabeled data \cite{Haresamudram2022_HAR_SSL_SOTA, Jain2022_ColloSSL, Matton2023_ContrastiveLearningStressDetection, Merrill2023_SSL, Deldari2022_BeyondJustVision, Deldari2023_LatentMasking, Saeed2021_SenseAndLearn}. Unlike supervised learning, where models rely on annotated and well-curated datasets, SSL leverages inherent structures within the data to create pseudo-labels, enabling the model to learn without explicit annotations. This approach is particularly valuable for wearables because collecting labeled data for diverse user activities and contexts can be challenging and time-consuming. SSL has been successfully applied across wearable applications including human activity recognition (HAR) \cite{Haresamudram2022_HAR_SSL_SOTA, Jain2022_ColloSSL}, stress detection \cite{Matton2023_ContrastiveLearningStressDetection}, flu, and COVID-19 prediction \cite{Merrill2023_SSL}. 

% \cite{Saeed2021_SenseAndLearn, Deldari2023_LatentMasking, Xu2021_LIMUBERT, Deldari2022_BeyondJustVision}

A prevalent SSL technique is masking, which refers to "\textit{masking out some of the observations in the time series}" \cite{Liu202_SSCL_Review},
% A prevalent SSL technique is masking, where certain parts of the input data are dropped, 
% requiring the model to reconstruct the masked data or infer human behavior from the masked data. 
to obtain generalizable feature representations for downstream tasks. Existing SSL masking approaches have mainly focused on a single modality, e.g., inertial measurement unit \cite{Xu2021_LIMUBERT}, reconstruct statistical features instead of the raw sensor data \cite{Saeed2021_SenseAndLearn}, and use models that do not capture the temporal aspect of the data \cite{Deldari2023_LatentMasking}. 
% providing a different aspect of the data that can be used in contrastive learning settings. 
% However, contrastive learning for wearable time series is still in the early stages of exploration \cite{Liu202_SSCL_Review}, showing the need for further investigation of standard and unified data augmentation techniques. 
While these approaches show that masking performs robust feature extraction, their applicability to diverse sensor data with different changing characteristics and with different temporal masking scales remains unclear. In addition, their comparison to conventional imputation approaches remains unknown. Indeed, SSL effectiveness in capturing temporal dependencies and patterns may vary across diverse sensor data, raising questions about its relevance for certain applications.  

In this work, we provide a comprehensive investigation of the effectiveness of masking for 10 wearable sensors with different changing frequencies and arbitrary masking lengths and compare it with traditional imputation strategies. 
% We explore their effectiveness in wearable time series data with different changing frequencies and arbitrary masking window lengths. 
% We focus on the scenario where wearable sensor data points are missing. 
We leverage transformer neural network architecture with the self-attention mechanism \cite{Vaswani2017_AttentionIsAllYouNeed}, which has shown outstanding performance in several domains \cite{Tashiro2021_CSDI, Horn2020_SetFunctionsForTimeSeries, Du2023_SAITS, Wu2020_AttentionBasedLearning, Phan2022_SleepTransformer, Lee2022_AutomaticSleepScoring, Yildiz2022_MultivariateTimeSeriesImputation}. 
% but has not previously been explored for wearable sensor data imputation. 
% We investigate the performance of the transformer on 10 data streams with different changing frequencies and arbitrary missing gaps. 
Further, we explore the impact of masked data imputation on downstream HAR and stress detection tasks using publicly available datasets and a patch-based transformer model. 

% Despite the prevalence of the self-attention mechanisms in several domains, their application for missing data imputation is under-explored. \citet{Yildiz2022_MultivariateTimeSeriesImputation}, for instance, ... 

% To evaluate the feasibility of our approach, we use an existing data set with 12 physiological parameters collected from 27 participants over one week. The dataset consists of sensor data collected using arm-worn devices in free-living environments. In particular, it contains heart rate, heart rate variability, blood perfusion, skin temperature, blood pulse wave, electrodermal activity, respiration rate, barometric pressure, accelerometer, and activity type. 
To perform the experiments, we use three existing datasets, namely, Novartis2020 \cite{Luo2020_AssessmentFatigue}, WESAD \cite{Schmidt2018_WESAD} and UCI-HAR \cite{Anguita2013_UCIHAR} that have been collected both in controlled and natural settings and contain a diverse range of time series, such as heart rate, skin temperature, accelerometer, and gyroscope.

% Contributions 
The main contributions of this paper are as follows:

\begin{itemize}
    % \item We compare state-of-the-art imputation techniques with deep learning to impute missing sensor data collected with wearable devices. The results show that our approach outperforms state-of-the-art techniques by 20-40\% for root mean squared error and 3-46\% for mean absolute error.
    \item We investigate the performance of the transformer model for reconstructing 10 types of physiological and behavioral wearable data. Our results show that, while simple techniques, such as linear interpolation are sufficient for data types that do not change frequently (e.g., skin temperature), the transformer is necessary for more complex signals  (e.g., heart rate). 
    \item We explore the reconstruction performance on arbitrary masking lengths. The results show that the transformer is more robust to longer missing sequences, outperforming baselines by 0.05-0.43 mean absolute error (MAE), 0.24-0.78 root mean squared error (RMSE), and with a higher correlation by 0.19-0.64. Common imputation strategies, however, are sufficient for imputing short windows.
    \item We test the performance of deep learning and traditional approaches on downstream tasks. Our results further confirm the need to use deep neural networks for imputing dynamically changing wearable time series collected in real-world scenarios and suggest using less complex strategies for short missing lengths. 
\end{itemize}

\section{Approach}
\begin{figure}[!tbp]
\centering
\includegraphics[scale=0.85]{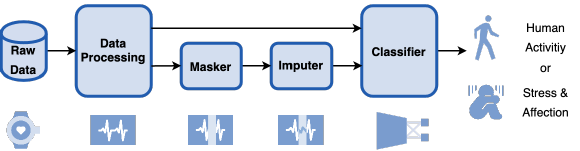}
\caption{Overview of our data analysis pipeline.}
\label{fig:pipeline}
\end{figure}

First, we selected three benchmark datasets and did the necessary preprocessing. We describe the datasets in detail in the Appendix. The processed datasets are then randomly masked to mimic real-world scenarios of missing data. Next, we trained a transformer model to reconstruct masked data and used the learned representations for the downstream classification tasks. Finally, we compared the classification results with training on the complete datasets. Figure \ref{fig:pipeline} provides an overview of our data analysis pipeline.

\textbf{Masked Data Reconstruction.} To train the model, we used the dataset collected by \cite{Luo2020_AssessmentFatigue}. We developed a deep neural network to predict raw missing sensor data collected in real-world scenarios. The network uses the transformer architecture, proposed by \citet{Vaswani2017_AttentionIsAllYouNeed}. Transformer is a sequence model, which has a main component known as \textit{self-attention mechanism} \cite{Wu2020_AttentionBasedLearning}. Sequence models such as, e.g., recurrent neural networks (RNNs), have commonly been used for time-series prediction, but they suffer from high computational complexity and vanishing gradient problems for long sequences \cite{Ribeiro2020_BeyondExploding}, making them less suitable for this task. The attention mechanism has become an effective alternative for such analysis thanks to its capability to capture dependencies of various ranges (e.g., shorter or longer) within a sequence. We used only the encoder of the transformer architecture because of its computational benefits 
% and combined it with a linear decoder \
similar to \cite{Yildiz2022_MultivariateTimeSeriesImputation}. The transformer operates over segments of sensor data that we expect to learn long-range dependencies in the input signal. The model takes as input a three-dimensional (3D) array or 3D tensor of shape (\textit{batch-size}, \textit{time-steps}, and \textit{channels}). The time steps refer to the length of the input sequence. Each input segment is first projected into a $d_{\textit{model}}$-dimensional space and summed with a learnable positional encoding matrix before feeding it into the transformer encoder. Positional encoding keeps track of the order of the sequence. We chose learnable over deterministic positional encodings as they have shown better performance in a similar setting \cite{Yildiz2022_MultivariateTimeSeriesImputation}. The encoder layer of the network is composed of a multi-head self-attention, a fully connected neural network, and residual connections around each of the sub-modules. We compare this model with traditional strategies such as \textit{mean}, \textit{median}, \textit{mode}, \textit{nearest neighbors}, \textit{linear interpolation}, and \textit{spline} \cite{Burkov2019_HundredPageML, Geron2019_HandsOnML2}.

\textbf{Downstream Classification Tasks.} To evaluate the performance in downstream tasks, we used UCI-HAR \cite{Anguita2013_UCIHAR} and WESAD \cite{Schmidt2018_WESAD} datasets, described in the Appendix. We defined three tasks: 6-class human activity classification for UCI HAR; \textit{stress} and \textit{non-stress} detection, and three-class classification into \textit{baseline}, \textit{stress} and \textit{amusement}. We then develop a neural network to predict these tasks using wearable data. Our model, based on the transformer architecture and influenced by the ViT network \cite{Dosovitskiy2020_ViT}, is tailored for handling multi-channel time series. 
% The model takes an input with dimension (sampling frequency in Hz X number of seconds, number of channels). 
The model employs the instance normalization layer to normalize each channel independently. Following this, the patch generation layer divides the time series window into shorter intervals. 
% resulting in X input patches with dimensions of X x Y. 
This process is reminiscent of tokenization in natural language processing tasks. This approach aids the transformer in detecting the parts of the signals that are crucial for activity and stress recognition. The patch embeddings and positional encodings are then provided as input to the transformer encoder. 
% with X blocks and X attention heads. 
The output of the model is a multi-layer perceptron. 
% with X and Y nodes. 
The model architecture is illustrated in Figure \ref{fig:VIT} in Appendix. 

% After the patches are generated, they are embedded into 64-dimensional vectors via a linear patch encoder layer. This is then added to 64-dimensional positional vectors to create images that encode both positional and waveform shape information of the input patches. The rest of the model is similar to a classic transformer encoder with 8 blocks with 4 attention heads, which is explained in more
% detail in Section A.2. Each block has a normalization layer, a multi-head attention layer, another
% normalization layer, and a two-layer multi-layer perceptron (MLP) with 128 and 64 units. For feature
% aggregation, we use global average pooling followed by a classification layer with units equal to the
% number of sleep stages, i.e. 5.

\section{Results}
In this section, we first discuss the results of masked data reconstruction for different data sources and masked window lengths as well as the performance in downstream tasks. 

\begin{table*}[htb]
\centering
\footnotesize
\caption{Mean and standard deviation of the mean absolute error ($MAE$), root mean squared error ($RMSE$) and Pearson correlation coefficient ($Pearson$) for each imputation strategy and each data source. We report the results after 100 runs.}
\label{table_imputation_results_sensor}
\begin{tabular}{ll|lllllll}
\toprule
\toprule
\textbf{Metric} & \textbf{Data}   &    \textbf{Linear} &           \textbf{Mean} &  \textbf{Median}   &          \textbf{Mode} &        \textbf{Nearest} &              \textbf{Spline} &    \textbf{Transformer}\\
             % &                &                &                &               &                        &                 &                \\
\midrule
 \textbf{MAE}   & \tt{ACC}    &  0.57 ± 0.04 &  0.57 ± 0.01 &  0.42 ± 0.02 &  \textbf{0.41 ± 0.02} &  0.58 ± 0.04 &  0.66 ± 0.39 &  \textbf{0.41 ± 0.01} \\
& \tt{BAR}         &  \textbf{0.21 ± 0.02} &  0.86 ± 0.02 &   0.84 ± 0.03 &  0.93 ± 0.04 &  \textbf{0.21 ± 0.02} &   0.55 ± 0.02 &  0.51 ± 0.02 \\
& \tt{BP}    &   \textbf{0.5 ± 0.02} &  0.84 ± 0.02 &  0.84 ± 0.02 &   1.1 ± 0.04 &   0.54 ± 0.02 &   0.73 ± 0.2 &  0.56 ± 0.01 \\
& \tt{BPW}    &  0.74 ± 0.02 &  0.76 ± 0.01 &  0.75 ± 0.01 &   1.29 ± 0.06 &    0.8 ± 0.03 &   0.75 ± 0.08 &  \textbf{0.62 ± 0.01} \\
& \tt{EE} &   0.47 ± 0.03 &  0.61 ± 0.02 &  0.43 ± 0.02 &  0.43 ± 0.02 &  0.48 ± 0.03 &  0.77 ± 0.89 &  \textbf{0.36 ± 0.01} \\
& \tt{HR}                &   0.47 ± 0.02 &  0.8 ± 0.02 &  0.8 ± 0.02 &  1.6 ± 0.08 &    0.51 ± 0.02 & 0.67 ± 0.02 &  \textbf{0.38 ± 0.01} \\
& \tt{HRV}               &   0.66 ± 0.02 &  0.82 ± 0.01 &  0.81 ± 0.02 &  1.12 ± 0.04 &  0.73 ± 0.02 &  0.73 ± 0.02 &  \textbf{0.65 ± 0.01} \\
& \tt{RESP}              &  0.79 ± 0.02 &  0.76 ± 0.01 &  0.74 ± 0.01 &  1.44 ± 0.09 &  0.88 ± 0.03 &  0.72 ± 0.04 &   \textbf{0.63 ± 0.01} \\
& \tt{ST}   &  \textbf{0.39 ± 0.02} &  0.82 ± 0.02 &  0.82 ± 0.02 &  0.94 ± 0.03 &   0.43 ± 0.02 &    0.77 ± 0.02 &  0.54 ± 0.02 \\
& \tt{Step}             &  0.36 ± 0.03 &  0.46 ± 0.01 &    \textbf{0.26 ± 0.02} &    \textbf{0.26 ± 0.02} &  0.36 ± 0.03 &  0.74 ± 1.02 &  0.32 ± 0.01 \\
\midrule
\textbf{RMSE}   & \tt{ACC}    &  1.32 ± 0.15 &  1.01 ± 0.03 &  1.08 ± 0.03 &  1.08 ± 0.03 &  1.47 ± 0.2 &    3.14 ± 7.17 &   \textbf{0.84 ± 0.03} \\
& \tt{BAR}         &   \textbf{0.51 ± 0.06} &  1.05 ± 0.03 &  1.13 ± 0.03 &  1.27 ± 0.04 &  0.58 ± 0.07 &    0.82 ± 0.06 &  0.72 ± 0.03 \\
& \tt{BP}    &  0.77 ± 0.06 &  1.04 ± 0.02 &   1.07 ± 0.03 &  1.45 ± 0.04 &  0.86 ± 0.06 &     1.42 ± 4.15 &  \textbf{0.75 ± 0.02} \\
& \tt{BPW}    &   1.08 ± 0.04 &   1.02 ± 0.02 &  1.03 ± 0.02 &  1.77 ± 0.07 &   1.2 ± 0.05 &   1.34 ± 1.38 &  \textbf{0.86 ± 0.01} \\
& \tt{EE} &  1.05 ± 0.07 &  1.02 ± 0.04 &  1.08 ± 0.04 &  1.09 ± 0.04 &  1.15 ± 0.09 &  3.78 ± 14.66 &  \textbf{0.64 ± 0.03} \\
& \tt{HR}                &  0.75 ± 0.03 &  1.04 ± 0.02 &  1.06 ± 0.03 &   2.09 ± 0.12 &  0.84 ± 0.04 &    0.92 ± 0.06 &   \textbf{0.53 ± 0.01} \\
& \tt{HRV}               &  0.92 ± 0.03 &  1.03 ± 0.02 &  1.04 ± 0.02 &  1.43 ± 0.04 &  1.01 ± 0.03 &     0.94 ± 0.03 &  \textbf{0.84 ± 0.01} \\
& \tt{RESP}               &  1.14 ± 0.04 &  1.02 ± 0.02 &  1.03 ± 0.02 &  2.11 ± 0.12 &  1.29 ± 0.05 &     1.06 ± 0.41 &  \textbf{0.87 ± 0.01} \\
& \tt{ST}   &  \textbf{ 0.69 ± 0.06} &  1.05 ± 0.02 &  1.07 ± 0.03 &  1.23 ± 0.03 &  0.76 ± 0.06 &    1.03 ± 0.05 &  0.74 ± 0.02 \\
& \tt{Step}             &  1.07 ± 0.08 &  0.98 ± 0.04 &  1.01 ± 0.04 &  1.01 ± 0.04 &  1.17 ± 0.1 &    6.16 ± 19.56 &  \textbf{0.74 ± 0.03} \\
\midrule
\textbf{Pearson}  & \tt{ACC}     &  0.22 ± 0.03 &  -0.17 ± 0.02 &  -0.03 ± 0.02 &  -0.0 ± 0.02 &  0.19 ± 0.03 &  0.09 ± 0.06 &  \textbf{0.54 ± 0.02} \\
& \tt{BAR}         &    \textbf{0.86 ± 0.03} &   -0.34 ± 0.03 &  -0.13 ± 0.04 &  -0.07 ± 0.04 &  0.83 ± 0.03 & 0.61 ± 0.05 &  0.69 ± 0.02 \\
& \tt{BP}    &   \textbf{0.68 ± 0.04} &  -0.3 ± 0.02 &  -0.17 ± 0.03 &  -0.04 ± 0.03 &  0.64 ± 0.03 &   0.41 ± 0.07 &  0.66 ± 0.01 \\
& \tt{BPW}    &  0.36 ± 0.03 &   -0.21 ± 0.02 &  -0.11 ± 0.02 &  -0.02 ± 0.02 &  0.32 ± 0.03 &  0.2 ± 0.07 & \textbf{0.59 ± 0.01} \\
& \tt{EE} &   0.39 ± 0.04 &  -0.21 ± 0.02 &   -0.02 ± 0.03 &  -0.0 ± 0.03 &  0.36 ± 0.04 &   0.1 ± 0.05 & \textbf{0.77 ± 0.01} \\
& \tt{HR}                &  0.7 ± 0.02 &  -0.29 ± 0.02 &  -0.19 ± 0.03 &  -0.0 ± 0.03 &  0.66 ± 0.02 &  0.46 ± 0.04 &  \textbf{0.85 ± 0.01} \\
& \tt{HRV}               &  \textbf{0.54 ± 0.02} &   -0.26 ± 0.02 &  -0.18 ± 0.02 &  -0.05 ± 0.03 &  0.51 ± 0.02 &   0.39 ± 0.03 &  \textbf{0.54 ± 0.01} \\
& \tt{RESP}              &  0.36 ± 0.02 &  -0.2 ± 0.02 &  -0.11 ± 0.02 &   0.0 ± 0.02c&  0.33 ± 0.02 &   0.26 ± 0.05 &  \textbf{0.50 ± 0.01} \\
& \tt{ST}   &  \textbf{0.75 ± 0.04} &  -0.31 ± 0.02 &  -0.18 ± 0.03 &  -0.06 ± 0.04 &  0.72 ± 0.04 &  0.27 ± 0.04 &   0.67 ± 0.02 \\
& \tt{Step}             &  0.28 ± 0.04 &  -0.18 ± 0.02 &     0.0 ± 0.02 &     0.0 ± 0.02 &  0.26 ± 0.04 &   0.05 ± 0.03 &  \textbf{0.65 ± 0.02} \\
\bottomrule
\bottomrule
\end{tabular}
\end{table*}

\begin{table*}[htb]
\centering
\footnotesize
\caption{Mean and standard deviation of the mean absolute error ($MAE$), root mean squared error ($RMSE$), and Pearson correlation coefficient ($Pearson$) for imputation strategies and window lengths (L, M, S). We report the results after 100 runs.}
\label{table_imputation_results_length}
\begin{tabular}{lllllllll}
\toprule
\toprule
& \textbf{Gap} &        \textbf{ Linear} &           \textbf{Mean} &         \textbf{Median} &          \textbf{ Mode} &       \textbf{ Nearest} &              \textbf{Spline} &   \textbf{Transformer} \\
& \textbf{}      &                &                &                &                &                &                                  &                \\
\midrule
\textbf{MAE}& \tt{L}    &  0.54 ± 0.01 &   0.72 ± 0.01 &  0.66 ± 0.01 &  0.92 ± 0.02 &  0.58 ± 0.01 &   0.66 ± 0.03 &  \textbf{0.49 ± 0.01} \\
& \tt{M} &   \textbf{0.47 ± 0.01} &  0.74 ± 0.01 &   0.68 ± 0.01 &  0.97 ± 0.02 &  0.50 ± 0.01 &   0.68 ± 0.18 &  0.49 ± 0.01 \\
& \tt{S}  &  \textbf{0.49 ± 0.02} &  0.85 ± 0.02 &   0.81 ± 0.02 &   1.15 ± 0.03 &  0.52 ± 0.02 &   0.75 ± 0.02 &  0.61 ± 0.01 \\
\midrule
\textbf{RMSE}& \tt{L}    &  0.98 ± 0.03 &   1.01 ± 0.01 &  1.04 ± 0.02 &  1.46 ± 0.04 &  1.09 ± 0.04 &    1.34 ± 1.53 &  \textbf{0.74 ± 0.01} \\
& \tt{M} &   0.91 ± 0.03 &  1.04 ± 0.01 &  1.07 ± 0.02 &  1.54 ± 0.03 &  1.01 ± 0.04 &  2.69 ± 10.1 &  \textbf{0.76 ± 0.01} \\
& \tt{S}   &  1.00 ± 0.05 &  1.24 ± 0.04 &   1.27 ± 0.04 &  1.8 ± 0.056 &  1.09 ± 0.05 &   1.19 ± 0.17 &  \textbf{0.99 ± 0.04} \\
\midrule
\textbf{Pearson}& \tt{L}    &   0.45 ± 0.02 &   -0.33 ± 0.01 &   -0.13 ± 0.02 &  -0.02 ± 0.02 &   0.42 ± 0.02 &   0.25 ± 0.07 &   \textbf{0.64 ± 0.01} \\
& \tt{M}  &  0.59 ± 0.02 &  -0.13 ± 0.02 &  -0.05 ± 0.01 &  -0.02 ± 0.01 &  0.55 ± 0.02 &    0.29 ± 0.11 &  \textbf{0.67 ± 0.01} \\
& \tt{S}  & \textbf{0.64 ± 0.03} &  -0.02 ± 0.03 &  -0.03 ± 0.02 &  -0.02 ± 0.02 &  0.60 ± 0.03 &   0.34 ± 0.05 &  0.59 ± 0.02 \\
\bottomrule
\bottomrule
\end{tabular}
\end{table*}

\begin{figure*}[htb]
\centering
 \begin{minipage}[b]{.48\textwidth}
    \centering
    \includegraphics[scale=0.4]{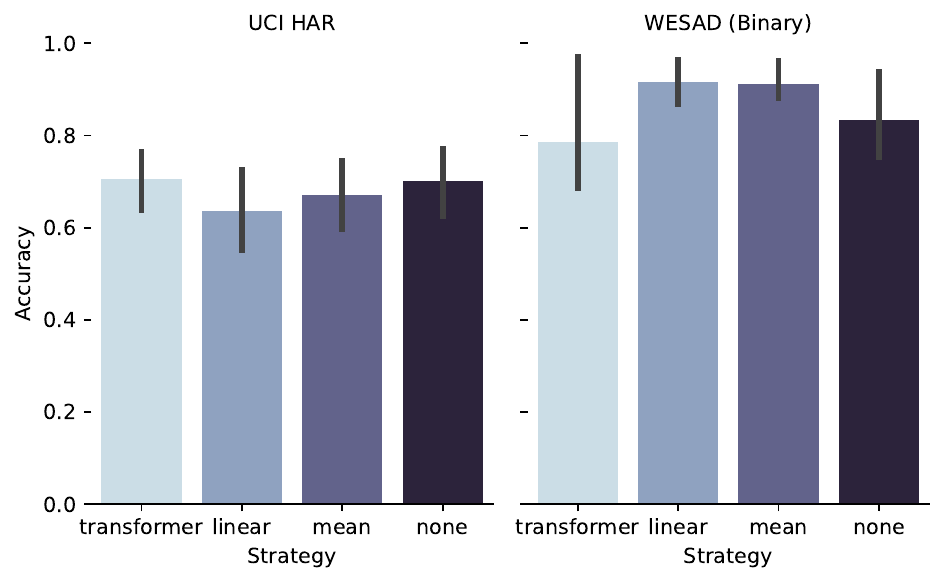}
    \caption{Performance of imputation strategies in downstream tasks: activity recognition and stress detection.}
    \label{fig:downstream_task_results_strategy}
 \end{minipage}
 \hspace{0.10pc}
 \begin{minipage}[b]{.48\textwidth}
    \centering
    \includegraphics[scale=0.4]{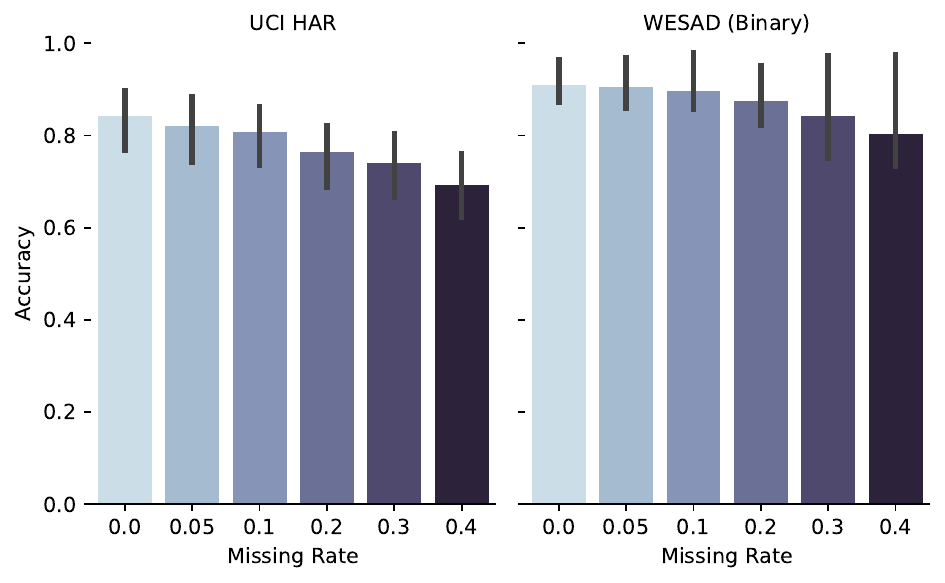}
    \caption{Performance in downstream tasks using the whole data set (missing data rate 0) and eliminating data (e.g., 0.1).}
    \label{fig:downstream_task_results_missingrate}
 \end{minipage}
\end{figure*}

% \subsection{Masked Data Reconstruction Performance}
% In this section, we discuss the results of the reconstruction by data source and masked data length.

\textbf{Masked Data Reconstruction Performance by Data Source.} 
% We then investigate the performance of our approach in comparison to the existing techniques for imputing missing data of a specific data stream. 
Table \ref{table_imputation_results_sensor} reports the MAE, RMSE, Pearson, and Spearman's rank correlation coefficients for each imputation strategy by the data source. Overall, the transformer model showed the highest capability to reconstruct masked data. However, it does not perform best for all data sources. In particular, it performs consistently better for dynamic data sources such as, e.g., accelerometer (ACC) (MAE=0.41), blood pulse wave (BPW) (MAE=0.62), energy expenditure (EE) (MAE=0.36), heart rate (HR) (MAE=0.38), heart rate variability (HRV) (MAE=0.65) and respiration rate (RESP) (MAE=0.63). However, this is not the case for more static data streams, such as the number of steps, skin temperature (ST), and barometer (BAR). For instance, linear interpolation performs better on BAR with an MAE of $0.21\pm0.02$ against the MAE of $0.51\pm0.02$ of the transformer. Linear interpolation further outperformed the transformer model on BP and ST data imputation. For the \textit{steps} variable median or mode imputation achieved the lowest MAE with $0.26\pm0.02$ and the highest Spearman's rank correlation coefficient with $0.69\pm0.03$. Spline and mean imputation did not perform best on any variable or performance metric. These results highlight the need to employ various imputation strategies for different data types depending on the signal's changing frequency. While some data sources (e.g., HR, HRV) require more sophisticated and complex reconstruction models, simpler techniques might be sufficient for other data sources (e.g., ST, steps). These results suggest employing hybrid reconstruction techniques in future work. 

\textbf{Masked Data Reconstruction Performance by Window Length.}
We then investigated the robustness of wearable data reconstruction depending on the length of the masked data window. Table \ref{table_imputation_results_length} presents the missing data prediction results using long (\textit{L}), medium (\textit{M}), and small (\textit{S}) masked sequence lengths. From the table, we observe that linear interpolation achieves an MAE of $0.49$ and a Pearson correlation of $0.64$ for imputing short sequences of data. The results are expected because in short physiological and behavioral sequences of data are unlikely to undergo large changes, and simply connecting the endpoints captures the underlying trend reasonably well. While the performance of linear interpolation is very good for short sequences, this performance drops significantly for large sequences. For medium to long sequences, the transformer achieves the lowest reconstruction error with an MAE of $0.49\pm0.01$, RMSE of $0.76\pm0.01$, and Pearson of $0.64\pm0.01$ on medium sequences and an MAE of $0.49\pm0.01$, RMSE of $0.74\pm0.01$ and Pearson of $0.67\pm0.01$ on long sequences. This implies that the transformer model is capable to learn representations of the signal and maintain its characteristics even in the case of long gaps. 

% \subsection{Downstream Classification Task Performance}
% In this section, we evaluate the impact of reconstruction strategies in downstream tasks during inference time. 

\textbf{Impact of Imputation on Downstream Tasks.} We then designed four sets of experiments to investigate the impact of reconstruction in downstream tasks: 1) \textit{none} refers to the well-curated dataset where all the data is present, which serves as the upper-bound performance; 2) \textit{mean} refers to replacing missing data with a mean value; 3) \textit{linear} replaces missing data by linearly interpolating the last observed sensor value with the next available value, and 4) \textit{transformer} refers to filling missing data with predictions of the transformer neural network. Figure \ref{fig:downstream_task_results_strategy} presents the accuracy of imputation strategies for human activity recognition using the UCI HAR dataset and stress detection using the WESAD dataset. We find that for UCI HAR the transformer network reaches an accuracy closer to the upper-bound performance in comparison to the other two imputation strategies. However, this is not the case for the WESAD dataset. This could be first because WESAD was collected in a controlled setting, which implies that the data is cleaner and more static than that of UCI HAR. Second, the data types in the two datasets are different. In particular, UCI HAR contains ACC and GYRO, which are more dynamically changing time series.

% Performance of the GB classifier, in terms of BA, using the whole
% data set (missing data rate = 0), and eliminating 10%, 25% and 50% of the
% sensor data points in a window using the Random technique.

\textbf{Impact of Missing Data Rate on Downstream Tasks.} We then investigated the impact of imputed data rate on the downstream task performance. Figure \ref{fig:downstream_task_results_missingrate} shows the classification accuracy of the transformer on the UCI HAR and WESAD datasets. We observe that for the UCI HAR dataset, the performance of the classifier with 40\% of imputed data is significantly lower in comparison to using all real data. The drop in performance is not as much for the WESAD dataset. This drop in performance for the UCI HAR could be because this dataset is more challenging to reconstruct correctly given that it was collected in a real-world setting.

\section{Conclusion}
In this work, we investigated the performance of imputation strategies for restoring missing sensor data collected with wearable devices.
% , which is often lost in multimodal, real-world data collection environments. 
To perform the study, we used three datasets collected both in controlled and real-world environments. 
% to train the imputers as well as evaluate their performance on downstream tasks. 
% arm-worn devices, which contains 12 physiological parameters of 27 participants collected over one week in free-living environments. 
% We explored different missing data imputation methods for wearable sensor data and demonstrated the feasibility of using a transformer model for signal reconstruction. 
% Our results show that the proposed method outperforms state-of-the-art imputation methods, such as e.g., linear interpolation and nearest neighbor imputation, by 20-40\% RMSE and 2-46\% MAE for reproducing the data from a lost modality. 
Our results show high stability of transformer-based imputation strategy for dynamically changing time-series data and for long and medium missing data sequence lengths. 
% We showed the advantages of hybrid imputation strategies that use different imputation methods based on the type of physiological variable and missing data length. 
Taken together, our results open up new opportunities for identifying pre-text tasks in SSL as well as the imputation strategy for pre-processing wearable data.

\textbf{Limitations and Future work.} While our study shows interesting insights related to the missing data problem for wearable devices, future work is needed to address some of the limitations of our work. An important limitation is that we relied on the data collected from a specific device. We are aware that there exist several wearable devices for different body positions capable of measuring similar sensor data. The device configuration (e.g., sampling rate) and placement could have an impact on overall findings due to the different movement patterns of different body parts. In future work, we plan to evaluate the approach in a cross-device setting with a different sampling frequency. 
% In the future, we can try improving preprocessing strategies and expanding the scope of datasets. 
% Firstly, given that the classification performance on WESAD data still needs much improvement, we may want to try manual normalization techniques on WESAD data (instead of / apart from learned normalization layers) to see if they can achieve more accurate results. 
In addition, we will incorporate a broader range of datasets, enabling a comprehensive evaluation on several types of wearable sensor data.
% In this project, we focused on evaluating imputation strategies while largely ignoring improving the overall performance of downstream tasks. So, in the future, we can push further into getting better classification results. For example, we can address class imbalance problems using strategies like SMOTE oversampling and random undersampling. 
In these experiments we investigated seven different imputation strategies, exploring other deep learning architectures to provide more insights into the strengths and weaknesses of different models is an interesting direction for future work.

% Use \bibliography{yourbibfile} instead or the References section will not appear in your paper
% \nobibliography{aaai22}
\bibliography{aaai22}

\begin{thebibliography}{39}
\providecommand{\natexlab}[1]{#1}

\bibitem[{Ad{\~a}o~Martins et~al.(2021)Ad{\~a}o~Martins, Annaheim, Spengler, and Rossi}]{Adao2021_FatigueMonitoring}
Ad{\~a}o~Martins, N.~R.; Annaheim, S.; Spengler, C.~M.; and Rossi, R.~M. 2021.
\newblock {Fatigue Monitoring Through Wearables: A State-Of-The-Art Review}.
\newblock \emph{Frontiers in physiology}, 2285.

\bibitem[{Allen(2007)}]{Allen2007_PPG}
Allen, J. 2007.
\newblock {Photoplethysmography and Its Application in Clinical Physiological Measurement}.
\newblock \emph{Physiological Measurement}, 28(3): R1.

\bibitem[{Anguita et~al.(2013)Anguita, Ghio, Oneto, Parra, Reyes-Ortiz et~al.}]{Anguita2013_UCIHAR}
Anguita, D.; Ghio, A.; Oneto, L.; Parra, X.; Reyes-Ortiz, J.~L.; et~al. 2013.
\newblock A public domain dataset for human activity recognition using smartphones.
\newblock In \emph{Esann}, volume~3, 3.

\bibitem[{Boucsein(2012)}]{Boucsein2012_EDA}
Boucsein, W. 2012.
\newblock \emph{{Electrodermal Activity}}.
\newblock Springer Science \& Business Media.

\bibitem[{Burkov(2019)}]{Burkov2019_HundredPageML}
Burkov, A. 2019.
\newblock \emph{{The Hundred-page Machine Learning Book}}, volume~1.
\newblock Andriy Burkov Quebec City, Can.

\bibitem[{Chen et~al.(2021)Chen, Zhang, Yao, Guo, Yu, and Liu}]{Chen2021_DL_HAR}
Chen, K.; Zhang, D.; Yao, L.; Guo, B.; Yu, Z.; and Liu, Y. 2021.
\newblock {Deep Learning for Sensor-Based Human Activity Recognition: Overview, Challenges, and Opportunities}.
\newblock \emph{ACM Computing Surveys (CSUR)}, 54(4): 1--40.

\bibitem[{Deldari et~al.(2023)Deldari, Spathis, Malekzadeh, Kawsar, Salim, and Mathur}]{Deldari2023_LatentMasking}
Deldari, S.; Spathis, D.; Malekzadeh, M.; Kawsar, F.; Salim, F.; and Mathur, A. 2023.
\newblock Latent Masking for Multimodal Self-supervised Learning in Health Timeseries.
\newblock \emph{arXiv preprint arXiv:2307.16847}.

\bibitem[{Deldari et~al.(2022)Deldari, Xue, Saeed, He, Smith, and Salim}]{Deldari2022_BeyondJustVision}
Deldari, S.; Xue, H.; Saeed, A.; He, J.; Smith, D.~V.; and Salim, F.~D. 2022.
\newblock Beyond just vision: A review on self-supervised representation learning on multimodal and temporal data.
\newblock \emph{arXiv preprint arXiv:2206.02353}.

\bibitem[{Dosovitskiy et~al.(2020)Dosovitskiy, Beyer, Kolesnikov, Weissenborn, Zhai, Unterthiner, Dehghani, Minderer, Heigold, Gelly et~al.}]{Dosovitskiy2020_ViT}
Dosovitskiy, A.; Beyer, L.; Kolesnikov, A.; Weissenborn, D.; Zhai, X.; Unterthiner, T.; Dehghani, M.; Minderer, M.; Heigold, G.; Gelly, S.; et~al. 2020.
\newblock An image is worth 16x16 words: Transformers for image recognition at scale.
\newblock \emph{arXiv preprint arXiv:2010.11929}.

\bibitem[{Du, C{\^o}t{\'e}, and Liu(2023)}]{Du2023_SAITS}
Du, W.; C{\^o}t{\'e}, D.; and Liu, Y. 2023.
\newblock {SAITS: Self-Attention-Based Imputation for Time Series}.
\newblock \emph{Expert Systems with Applications}, 119619.

\bibitem[{Geron(2019)}]{Geron2019_HandsOnML2}
Geron, A. 2019.
\newblock \emph{{Hands-On Machine Learning with Scikit-Learn, Keras, and TensorFlow: Concepts, Tools, and Techniques to Build Intelligent Systems}}.
\newblock O'Reilly Media.

\bibitem[{Guan and Pl{\"o}tz(2017)}]{Guan2017_EnsemblesDeepLSTMHAR}
Guan, Y.; and Pl{\"o}tz, T. 2017.
\newblock {Ensembles of Deep LSTM Learners for Activity Recognition using Wearables}.
\newblock \emph{Proceedings of the ACM on Interactive, Mobile, Wearable and Ubiquitous Technologies (PACM IMWUT)}, 1(2): 1--28.

\bibitem[{Haresamudram, Essa, and Pl{\"o}tz(2022)}]{Haresamudram2022_HAR_SSL_SOTA}
Haresamudram, H.; Essa, I.; and Pl{\"o}tz, T. 2022.
\newblock Assessing the state of self-supervised human activity recognition using wearables.
\newblock \emph{Proceedings of the ACM on Interactive, Mobile, Wearable and Ubiquitous Technologies}, 6(3): 1--47.

\bibitem[{Heaney(2020)}]{Heaney2020_EE}
Heaney, J. 2020.
\newblock {Energy: Expenditure, Intake, Lack of}.
\newblock In \emph{Encyclopedia of Behavioral Medicine}, 778--779. Springer.

\bibitem[{Horn et~al.(2020)Horn, Moor, Bock, Rieck, and Borgwardt}]{Horn2020_SetFunctionsForTimeSeries}
Horn, M.; Moor, M.; Bock, C.; Rieck, B.; and Borgwardt, K. 2020.
\newblock {Set Functions for Time Series}.
\newblock In \emph{International Conference on Machine Learning (ICML)}, 4353--4363. PMLR.

\bibitem[{Jain et~al.(2022)Jain, Tang, Min, Kawsar, and Mathur}]{Jain2022_ColloSSL}
Jain, Y.; Tang, C.~I.; Min, C.; Kawsar, F.; and Mathur, A. 2022.
\newblock Collossl: Collaborative self-supervised learning for human activity recognition.
\newblock \emph{Proceedings of the ACM on Interactive, Mobile, Wearable and Ubiquitous Technologies}, 6(1): 1--28.

\bibitem[{Lee and Saeed(2018)}]{Lee2022_AutomaticSleepScoring}
Lee, H.; and Saeed, A. 2018.
\newblock {Automatic Sleep Scoring from Large-scale Multi-channel Pediatric EEG}.
\newblock In \emph{NeurIPS 2022 Workshop on Learning from Time Series for Health}. IEEE.

\bibitem[{Liu et~al.(2023)Liu, Alavi, Li, and Zhang}]{Liu202_SSCL_Review}
Liu, Z.; Alavi, A.; Li, M.; and Zhang, X. 2023.
\newblock Self-Supervised Contrastive Learning for Medical Time Series: A Systematic Review.
\newblock \emph{Sensors}, 23(9): 4221.

\bibitem[{Luo et~al.(2020)Luo, Lee, Clay, Jaggi, and De~Luca}]{Luo2020_AssessmentFatigue}
Luo, H.; Lee, P.-A.; Clay, I.; Jaggi, M.; and De~Luca, V. 2020.
\newblock {Assessment of Fatigue Using Wearable Sensors: A Pilot Study}.
\newblock \emph{Digital biomarkers}, 4(1): 59--72.

\bibitem[{Marvin(1912)}]{Marvin1912_Barometers}
Marvin, C.~F. 1912.
\newblock \emph{{Barometers and the Measurement of Atmospheric Pressure: A Pamphlet of Information Respecting the Theory and Construction of Barometers in General, with Summary of Instructions for the Care and Use of the Standard Weather Bureau Instruments}}.
\newblock 472. US Government Printing Office.

\bibitem[{Matton et~al.(2023)Matton, Lewis, Guttag, and Picard}]{Matton2023_ContrastiveLearningStressDetection}
Matton, K.; Lewis, R.; Guttag, J.; and Picard, R. 2023.
\newblock Contrastive Learning of Electrodermal Activity Representations for Stress Detection.
\newblock In \emph{Conference on Health, Inference, and Learning}, 410--426. PMLR.

\bibitem[{Merrill and Althoff(2023)}]{Merrill2023_SSL}
Merrill, M.~A.; and Althoff, T. 2023.
\newblock Self-Supervised Pretraining and Transfer Learning Enable Flu and COVID-19 Predictions in Small Mobile Sensing Datasets.
\newblock In \emph{Conference on Health, Inference, and Learning}, 191--206. PMLR.

\bibitem[{Ord{\'o}{\~n}ez and Roggen(2016)}]{Ordonez2016_DeepConvLSTMHAR}
Ord{\'o}{\~n}ez, F.~J.; and Roggen, D. 2016.
\newblock {Deep Convolutional and LSTM Recurrent Neural Networks for Multimodal Wearable Activity Recognition}.
\newblock \emph{Sensors}, 16(1): 115.

\bibitem[{Phan et~al.(2022)Phan, Mikkelsen, Ch{\'e}n, Koch, Mertins, and De~Vos}]{Phan2022_SleepTransformer}
Phan, H.; Mikkelsen, K.; Ch{\'e}n, O.~Y.; Koch, P.; Mertins, A.; and De~Vos, M. 2022.
\newblock {SleepTransformer: Automatic Sleep Staging with Interpretability and Uncertainty Quantification}.
\newblock \emph{IEEE Transactions on Biomedical Engineering}, 69(8): 2456--2467.

\bibitem[{Picard(2000)}]{Picard2000_AffectiveComputing}
Picard, R.~W. 2000.
\newblock \emph{{Affective Computing}}.
\newblock MIT press.

\bibitem[{Pl{\"o}tz(2021)}]{Plotz2021ApplyingMLforSensorData}
Pl{\"o}tz, T. 2021.
\newblock {Applying Machine Learning for Sensor Data Analysis in Interactive Systems: Common Pitfalls of Pragmatic Use and Ways to Avoid Them}.
\newblock \emph{ACM Computing Surveys (CSUR)}, 54(6): 1--25.

\bibitem[{Radu et~al.(2018)Radu, Tong, Bhattacharya, Lane, Mascolo, Marina, and Kawsar}]{Radu2018_MultimodalDL}
Radu, V.; Tong, C.; Bhattacharya, S.; Lane, N.~D.; Mascolo, C.; Marina, M.~K.; and Kawsar, F. 2018.
\newblock {Multimodal Deep Learning for Activity and Context Recognition}.
\newblock \emph{Proceedings of the ACM on Interactive, Mobile, Wearable and Ubiquitous Technologies (PACM IMWUT)}, 1(4): 1--27.

\bibitem[{Ribeiro et~al.(2020)Ribeiro, Tiels, Aguirre, and Sch{\"o}n}]{Ribeiro2020_BeyondExploding}
Ribeiro, A.~H.; Tiels, K.; Aguirre, L.~A.; and Sch{\"o}n, T. 2020.
\newblock {Beyond Exploding and Vanishing Gradients: Analysing RNN Training Using Attractors and Smoothness}.
\newblock In \emph{International Conference on Artificial Intelligence and Statistics}, 2370--2380. PMLR.

\bibitem[{Saeed, Ungureanu, and Gfeller(2021)}]{Saeed2021_SenseAndLearn}
Saeed, A.; Ungureanu, V.; and Gfeller, B. 2021.
\newblock Sense and learn: Self-supervision for omnipresent sensors.
\newblock \emph{Machine Learning with Applications}, 6: 100152.

\bibitem[{Sano and Picard(2013)}]{Sano2013_StressDetection}
Sano, A.; and Picard, R.~W. 2013.
\newblock {Stress Recognition using Wearable Sensors and Mobile Phones}.
\newblock In \emph{2013 Humaine Association Conference on Affective Computing and Intelligent Interaction}, 671--676. IEEE.

\bibitem[{Schmidt et~al.(2018)Schmidt, Reiss, Duerichen, Marberger, and Van~Laerhoven}]{Schmidt2018_WESAD}
Schmidt, P.; Reiss, A.; Duerichen, R.; Marberger, C.; and Van~Laerhoven, K. 2018.
\newblock Introducing , a multimodal dataset for wearable stress and affect detection.
\newblock In \emph{Proceedings of the 20th ACM international conference on multimodal interaction}, 400--408.

\bibitem[{Tashiro et~al.(2021)Tashiro, Song, Song, and Ermon}]{Tashiro2021_CSDI}
Tashiro, Y.; Song, J.; Song, Y.; and Ermon, S. 2021.
\newblock {CSDI: Conditional Score-based Diffusion Models for Probabilistic Time Series Imputation}.
\newblock \emph{Advances in Neural Information Processing Systems}, 34: 24804--24816.

\bibitem[{Tong et~al.(2019)Tong, Craner, Vegreville, and Lane}]{Tong2019_TrackingFatigueMS}
Tong, C.; Craner, M.; Vegreville, M.; and Lane, N.~D. 2019.
\newblock {Tracking Fatigue and Health State in Multiple Sclerosis Patients Using Connnected Wellness Devices}.
\newblock \emph{Proceedings of the ACM on Interactive, Mobile, Wearable and Ubiquitous Technologies (PACM IMWUT)}, 3(3): 1--19.

\bibitem[{Vaswani et~al.(2017)Vaswani, Shazeer, Parmar, Uszkoreit, Jones, Gomez, Kaiser, and Polosukhin}]{Vaswani2017_AttentionIsAllYouNeed}
Vaswani, A.; Shazeer, N.; Parmar, N.; Uszkoreit, J.; Jones, L.; Gomez, A.~N.; Kaiser, {\L}.; and Polosukhin, I. 2017.
\newblock {Attention Is All You Need}.
\newblock \emph{Advances in neural information processing systems}, 30.

\bibitem[{Wu et~al.(2020)Wu, Zhang, Ilyas, and Rekatsinas}]{Wu2020_AttentionBasedLearning}
Wu, R.; Zhang, A.; Ilyas, I.; and Rekatsinas, T. 2020.
\newblock {Attention-Based Learning for Missing Data Imputation in HoloClean}.
\newblock \emph{Proceedings of Machine Learning and Systems}, 2: 307--325.

\bibitem[{Xu et~al.(2021)Xu, Zhou, Tan, Li, and Shen}]{Xu2021_LIMUBERT}
Xu, H.; Zhou, P.; Tan, R.; Li, M.; and Shen, G. 2021.
\newblock {LIMU-BERT: Unleashing the potential of unlabeled data for imu sensing applications}.
\newblock In \emph{Proceedings of the 19th ACM Conference on Embedded Networked Sensor Systems}, 220--233.

\bibitem[{Y{\i}ld{\i}z, Ko{\c{c}}, and Ko{\c{c}}(2022)}]{Yildiz2022_MultivariateTimeSeriesImputation}
Y{\i}ld{\i}z, A.~Y.; Ko{\c{c}}, E.; and Ko{\c{c}}, A. 2022.
\newblock Multivariate Time Series Imputation With Transformers.
\newblock \emph{IEEE Signal Processing Letters}, 29: 2517--2521.

\bibitem[{Zhai et~al.(2021)Zhai, Guan, Catt, and Pl{\"o}tz}]{Zhai2021_UbiSleepNet}
Zhai, B.; Guan, Y.; Catt, M.; and Pl{\"o}tz, T. 2021.
\newblock {Ubi-SleepNet: Advanced Multimodal Fusion Techniques for Three-stage Sleep Classification Using Ubiquitous Sensing}.
\newblock \emph{Proceedings of the ACM on Interactive, Mobile, Wearable and Ubiquitous Technologies (PACM IMWUT)}, 5(4): 1--33.

\bibitem[{Zhai et~al.(2020)Zhai, Perez-Pozuelo, Clifton, Palotti, and Guan}]{Zhai2020_MakingSenseOfSleep}
Zhai, B.; Perez-Pozuelo, I.; Clifton, E.~A.; Palotti, J.; and Guan, Y. 2020.
\newblock {Making Sense of Sleep: Multimodal Sleep Stage Classification in a Large, Diverse Population Using Movement and Cardiac Sensing}.
\newblock \emph{Proceedings of the ACM on Interactive, Mobile, Wearable and Ubiquitous Technologies (PACM IMWUT)}, 4(2): 1--33.

\end{thebibliography}

\section{Appendices.}
% Any appendices must appear after the main content. If your main sections are numbered, appendix sections must use letters instead of arabic numerals. In \LaTeX{} you can use the \texttt{\textbackslash appendix} command to achieve this effect and then use \texttt{\textbackslash section\{Heading\}} normally for your appendix sections.

\subsection{Datasets} \label{section_datasets}

\textbf{Novartis2020 \cite{Luo2020_AssessmentFatigue}.} The dataset contains \textit{wearable sensor} data collected from 28 participants over one week in free-living conditions. Participants are between 26 and 55 years old (Mean=42), 57\% are males, 39\% females, and the remaining of non-specified gender. Participants were asked to continuously wear a multisensor wearable device during the study to collect sensor data and use a mobile application to report daily fatigue questionnaires. To train the deep neural network, we use only the wearable sensor data, which was collected using the Everion Biovotion\footnote{Currently, acquired by Bifourmis \url{https://biofourmis.com/}.} wearable device. Participants were asked to wear the device on their non-dominant arm to reduce the presence of motion artifacts. Everion Biovotion contains five sensors, including, a three-axis accelerometer ({\tt ACC}) \cite{Chen2021_DL_HAR}, barometer ({\tt BAR}) \cite{Marvin1912_Barometers}, galvanic skin response ({\tt GSR}) \cite{Boucsein2012_EDA}, temperature ({\tt TEMP}) and photoplethysmography ({\tt PPG}) \cite{Allen2007_PPG}. Such sensors have been used to derive 12 physiological parameters at 1Hz sampling frequency. These parameters include participants' \textit{heart rate (HR)} -- the number of times the heart beats per minute \cite{Allen2007_PPG}--, \textit{heart rate variability (HRV)} -- the time variation between two heartbeats \cite{Allen2007_PPG} --, \textit{respiratory rate (RESP)} -- the number of breaths a person takes per minute \cite{Allen2007_PPG} --, \textit{blood perfusion} -- the process of a body delivering blood to a capillary bed in its biological tissue \cite{Allen2007_PPG} --, \textit{blood pulse wave (BPW)} -- when the heart contracts, blood is ejected generating a pulse wave that travels through the circulatory system, this metric measures both the shape and rhythmicity of the wave--, \textit{galvanic skin response (GSR)} -- refers to the arousal of the sympathetic nervous system \cite{Boucsein2012_EDA}--, \textit{activity counts} -- the intensity of movement in three-axis \cite{Chen2021_DL_HAR} --, \textit{activity class} -- resting, walking, running, cycling, biking and more --, \textit{steps} -- number of steps performed while walking --, \textit{energy expenditure (EE)} -- the amount of energy a person uses to complete bodily functions, from moving, breathing, digestion, respiration \cite{Heaney2020_EE} --, \textit{skin temperature} -- human body temperature --, and \textit{barometer} -- altitude changes of the user while wearing the device --. The physiological data is available at a one-minute level, i.e., one measurement per minute. There are in total 545 days with sensor data in the dataset. The dataset is publicly available for download\footnote{\url{https://zenodo.org/record/4266157}}.

\textbf{WESAD \cite{Schmidt2018_WESAD}.} WESAD is a multimodal wearable stress and affect detection dataset featuring physiological and motion data recorded from 15 subjects. It consists of both wirst-worn and chest-worn device data. Here we only use the signals from the Empatica E4 wrist-worn device, including blood volume pulse (BVP), sampled at 64 Hz, electrodermal activity (EDA), sampled at 4 Hz, body temperature (TEMP) sampled at 4 Hz, and triaxial acceleration (ACC) sampled at 32 Hz. Sensor data was collected while participants were exposed to tasks that cause \textit{stress}, \textit{amusemenet}, or \textit{baseline}, which we use to label sensor data. To preprocess the data, we first filtered the EDA signal with a second-order Butterworth low-pass filter with a cutoff frequency of 0.5 Hz. Then, we downsampled the signals to 4 Hz, the minimum sampling rate of all channels. Finally, we sliced the data into one-minute windows (240 readings/window) with 0.25s overlap (one reading) unless a different label occurs in a window. In this case, the starting point of the next window is shifted to where the first reading with the different label is located. This ensures that every window has consistent labels for all its readings. The readings before the new starting point are discarded as they are insufficient for a full window. Then we down-sampled the signal to a 4 Hz frequency to synchronize all time series. We then segmented the signals using 1-minute windows with a 0.25-second overlap. To label sensor data, we used the baseline, stress, and amusement parts of the protocol. We split the dataset into train and test sets, with 11 subjects for training (1030 window slices) and the rest 4 for testing (372 window slices).

\textbf{UCI-HAR \cite{Anguita2013_UCIHAR}.} UCI HAR is a dataset collected for enabling human activity recognition database. It consists of three-axis accelerometer and three-axis gyroscope data captured with a sampling rate of 50 Hz. UCI-HAR contains data from 30 participants performing 6 physical activities, such as, \textit{walking}, \textit{walking upstairs}, \textit{walking downstairs}, \textit{sitting}, \textit{standing} and \textit{laying down}. Participants performed the activities while carrying a waist-mounted smartphone with embedded inertial sensors. To preprocess the data, we applied median and 3rd-order low-pass Butterworth filters with a 20 Hz cutoff frequency to reduce the noise. We then removed the gravitational components from the raw accelerometer signals using a Butterworth low-pass filter with a 0.3 Hz cutoff frequency, leaving only the body motion acceleration parts. Finally, the resulting signals are sampled in fixed-width sliding windows of 2.56 seconds (128 readings/window) with 50\% overlap (64 readings). We used the 6 classes mentioned before to label the sensor data. We randomly partitioned the dataset into train and test sets by selecting 70\% participants for the training set (21 subjects, 7352 window slices) and 30\% for the test set (9 subjects, 2947 window slices).

\subsection{Missing Data Problem}

\begin{figure*}[!tbp]
\centering
 \begin{minipage}[b]{.48\textwidth}
    \centering
    \includegraphics[scale=0.57]{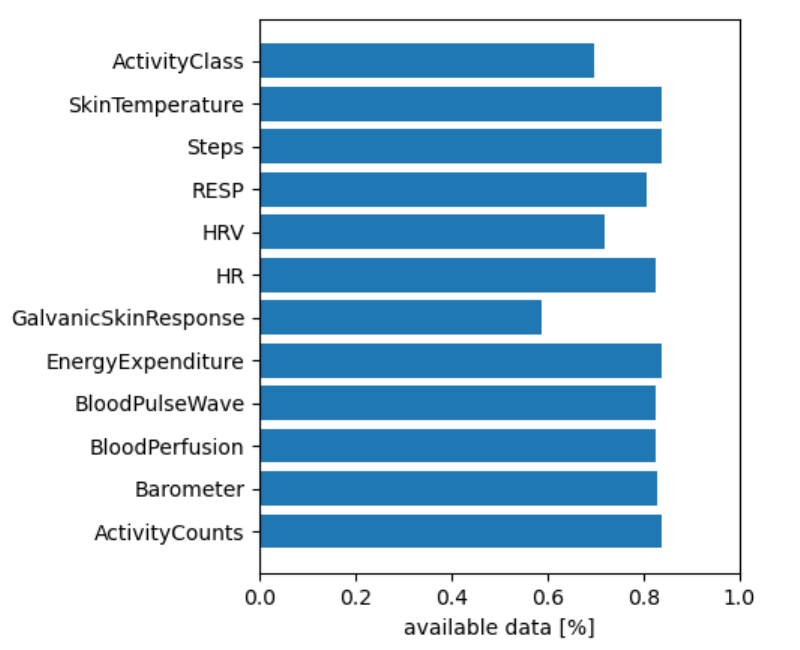}
    \caption{Overview of the available data per physiological variable. The x-axis shows the amount of data in percentage (\%) and the y-axis shows each physiological parameter.}
    \label{fig:missing_data_per_variable}
 \end{minipage}
 \hspace{0.01pc}
 \begin{minipage}[b]{.48\textwidth}
    \centering
    \includegraphics[scale=0.25]{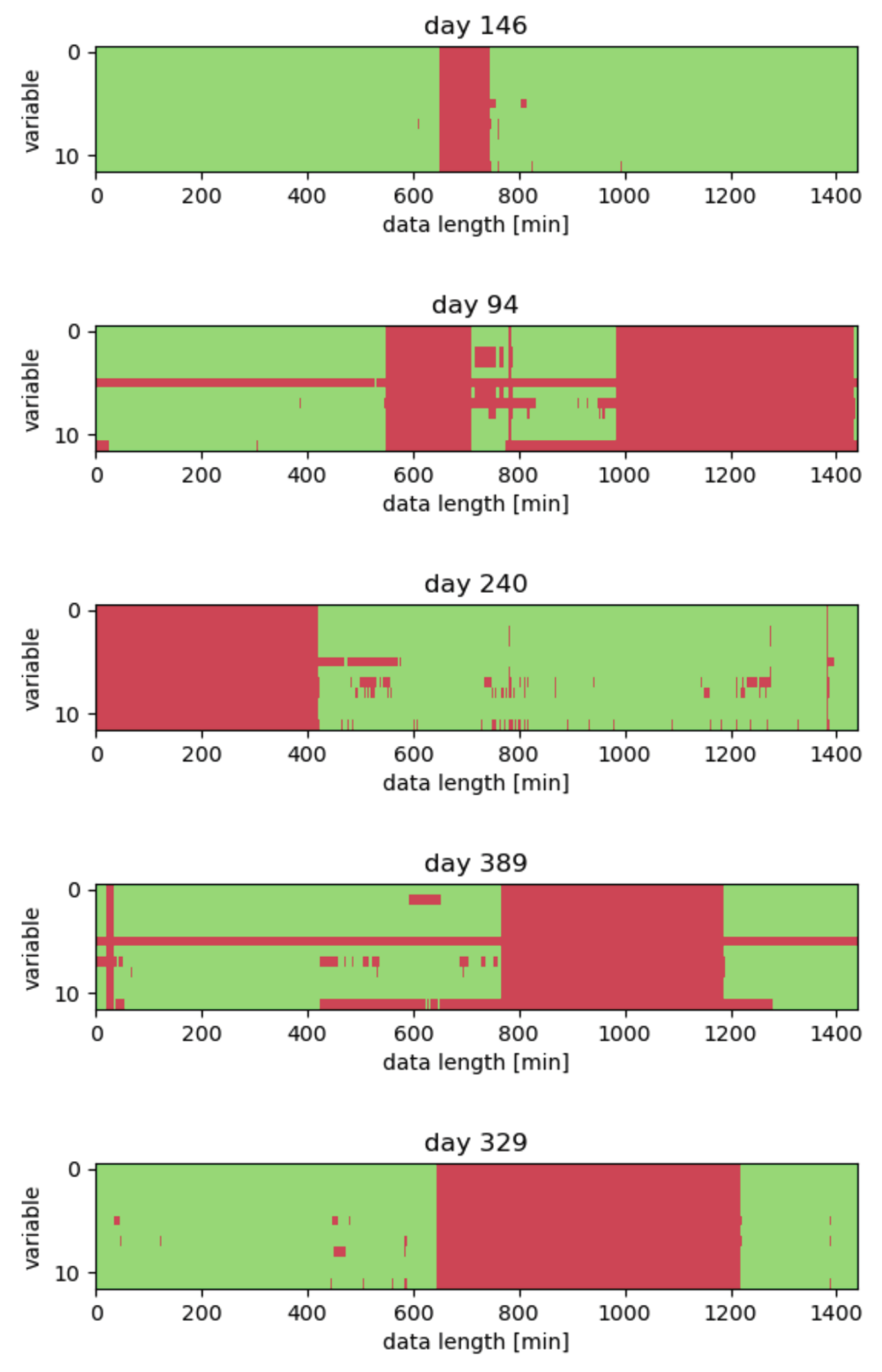}
    \caption{Overview of the missing data patterns for an exemplary set of days. The x-axis shows time in minutes and the y-axis shows the 10 physiological parameters considered in this work.}
    \label{fig:missing_data_example}
 \end{minipage}
\end{figure*}

To obtain a general understanding of the amount of missing data in the dataset, we first calculate the amount of missing data by parameter. Figure \ref{fig:missing_data_per_variable} shows the percentage of missing data for each physiological parameter. From the figure, we observe that overall parameters derived from TEMP, ACC, BAR, and PPG sensors have a similar amount of missing data (around 20\%), with an exception of the activity class and HRV parameters, for which the amount of missingness is higher, approximately 30\%. The sensor with the highest amount of missing data is EDA (approximately 45\%).

We then explore the patterns of missing data by visualizing the data for each physiological parameter for a day. Figure \ref{fig:missing_data_example} presents several example days of data collected from different participants in the form of a heat map. The x-axis represents the time of the day in minutes and each heat map row shows whether the data point for that physiological variable is either missing (red) or present (green). From the figure, we observe that the data can either be missing all at once -- meaning that all physiological parameters are missing -- or individually by sensor type. For instance, the third heat map (day 240) shows missing data from all sensors at once, which could be due to the user not wearing the device (i.e., due to charging or forgetting). Discarding all the days of data with similar missingness patterns would lead to a significant reduction of the dataset. For this reason, a sophisticated imputation method is crucial to fill the gaps and be able to utilize the available data.

% \begin{figure*}
%     \centering
%     \includegraphics[scale=0.25]{Figures/intra-subject_variability.png}
%     \caption{Distribution of the physiological and behavioral signals explored in this work. The x-axis presents the subject identifier and the y-axis the sensor value. From the figure, it is possible to observe the intra-subject variability by the data source.}
%     \label{fig:intra_subj_var}
% \end{figure*}

\subsection{Approach}

\subsubsection{Missing Data Imputation Model.} The encoder layer of the network is composed of a 1) multi-head self-attention, 2) a fully-connected neural network, and 3) residual connections around each of the sub-modules. Multi-head self-attention employs the self-attention mechanism, which for a given input sequence with multiple channels \( \mathcal{X \in \mathbb{R}^{T \times C}} \) of length \( T\) and the number of channels \( C\), is formulated as follows:

\[
Attention(Q, K, V) = Softmax(\frac{QK^T}{\sqrt{d_k}})V
\]
where the \( Q, K, V\) represent the query, key, and value vectors, respectively; \( d_{k}\) is the dimension of the keys. This mechanism allows the model to associate each sensor data point in the input with other data points. To achieve self-attention, the input is fed into three fully connected layers to create the query, key, and value vectors. The \( Q, K, V\) are then split into \( H\) vectors, known as \textit{heads} and each head performs the self-attention mechanism individually, formulated as follows:
% The network has then a multi-head operation, which performs the self-attention mechanism multiple times. 
\[
MultiHead(Q, K, V) = Concat(h_1, h_2, ..., h_H)W^O
\]

\[
h_i = Attention(Q W^Q_i, K W^K_i, V W^V_i)
\]
where \( W^Q_i, W^K_i, W^V_i\) are linear projection parameter matrices for the i-th head; \( H \) is the output of the multi-head self-attention operation. We employed four heads (\( H=4\)) similar to the initially proposed architecture in \cite{Vaswani2017_AttentionIsAllYouNeed}. The multi-head operation allows the model to jointly attend to information from different representations at different positions. 
% The encoder consists of three layers, each with multi-head self-attention (in our case four heads), similar to the initially proposed architecture in \cite{Vaswani2017_AttentionIsAllYouNeed}.
This layer is followed by a fully-connected layer with a Gaussian Error Linear Unit (GELU) with 18 hidden units and a batch-normalization layer.
% instead of the commonly used normalization layer in natural language processing (NLP) \cite{Vaswani2017_AttentionIsAllYouNeed}. 
% Additionally, we replace the rectified linear unit (ReLU) activation function with the Gaussian Error Linear Unit (GELU) in the feed-forward network. 
We use the same dimension for all elements in the self-attention layers (i.e. query, key, value weight matrices), which is $d_q = d_k = d_v = d_{\textit{model}}=16$.  We train the model using a learning rate of $1e-3$ for 400 epochs. Due to the efficient design of this architecture \cite{Yildiz2022_MultivariateTimeSeriesImputation} the model contains only 5434 learnable parameters, which could run efficiently on wearable devices, making the proposed approach feasible for a real scenario.

\subsubsection{Downstream Classification Model.}
The hyperparameters during the training include:

\begin{itemize}
    \item patch size ($P$): the size of the input patch sliced from a window
    \item depth ($D$): the number of blocks in the Transformer encoder
    \item number of heads ($H$): the number of attention heads in one block of the Transformer encoder
    \item embedding dimension ($d_{\mathrm{emb}}$): the dimension of the patch embeddings
    \item head dimension ($d_{\mathrm{attn}}$): the output dimension of the key, query, and value matrices in one attention head
    \item MLP dimension ($d_{\mathrm{mlp}}$): the dimension of the hidden layer of the MLP layers in the Transformer encoder
    \item embedding dropout ($p_{\mathrm{emb}}$): the dropout rate of the input embeddings
    \item attention dropout ($p_{\mathrm{attn}}$): the dropout rate in one block of the Transformer encoder
    \item MLP dropout ($p_{\mathrm{mlp}}$): the dropout rate of the MLP layers in the Transformer encoder.
\end{itemize}

Note that the MLP head for classification only contains a single layer. We referred to another paper\cite{Lee2022_AutomaticSleepScoring} for setting up some model parameters. As a result, we use $P=16$, $D=8$, $H=4$, $d_{\mathrm{emb}}=d_{\mathrm{attn}}=64$, $d_{\mathrm{mlp}}=128$, and $p_{\mathrm{emb}}=p_{\mathrm{attn}}=p_{\mathrm{mlp}}=0.4$, resulting in 667K parameters in total. This is suitable for operating on wearable devices. To reduce overfitting and evaluate the model's performance on different subjects, we used Leave-One-Subject-Out Cross-Validation during training for model selection, i.e., we left out all the data from one subject as the validation set for each cross-validation fold.

\begin{figure}
    \centering
    \includegraphics[scale=0.2]{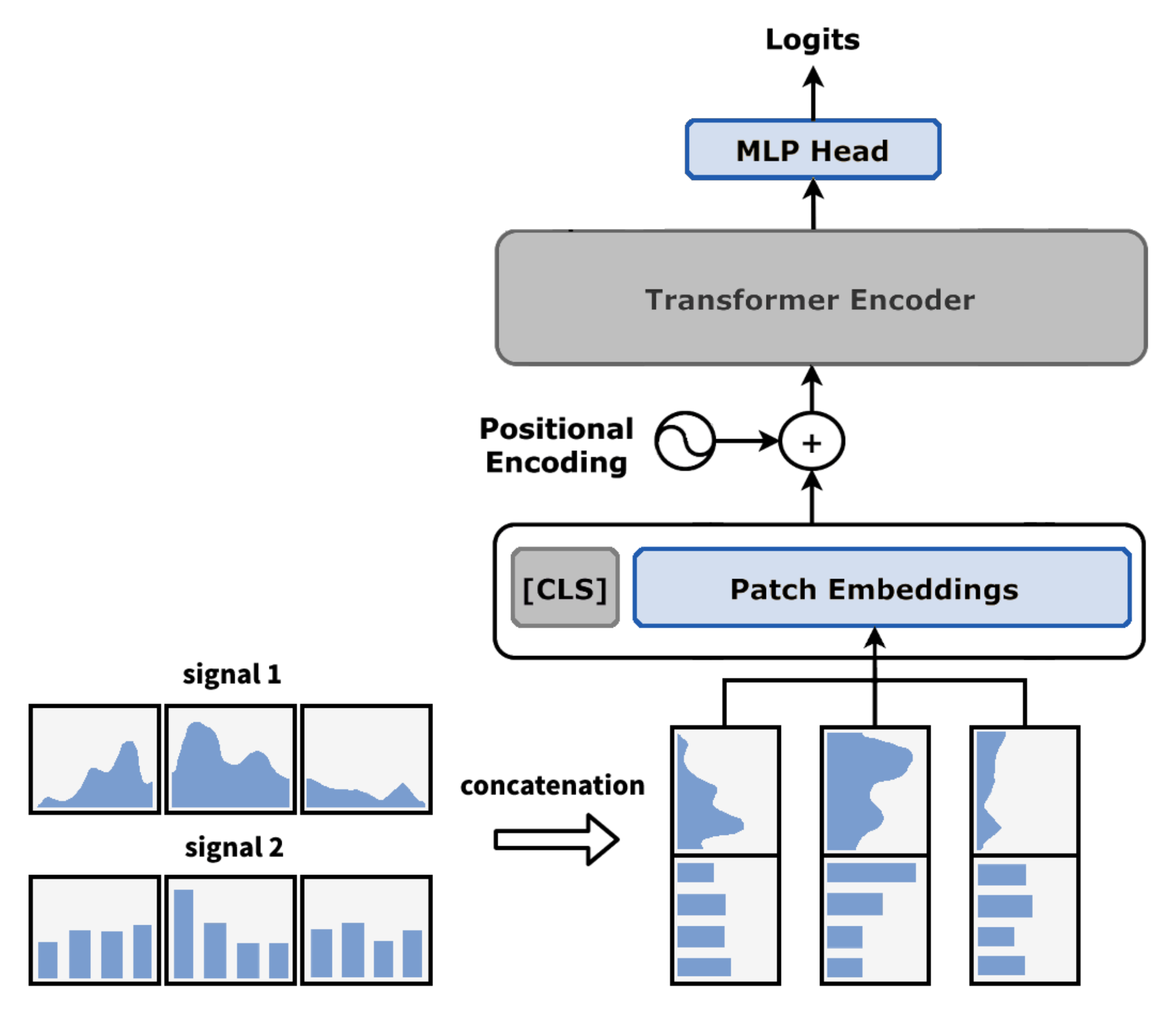}
    \caption{Overview of the patch-based transformer neural network architecture we used for downstream classification tasks. }
    \label{fig:VIT}
\end{figure}

\subsection{Results}
\begin{figure*}[!tbp]
\centering
\begin{minipage}[b]{.44\textwidth}
      \centering
    \includegraphics[scale=0.3]{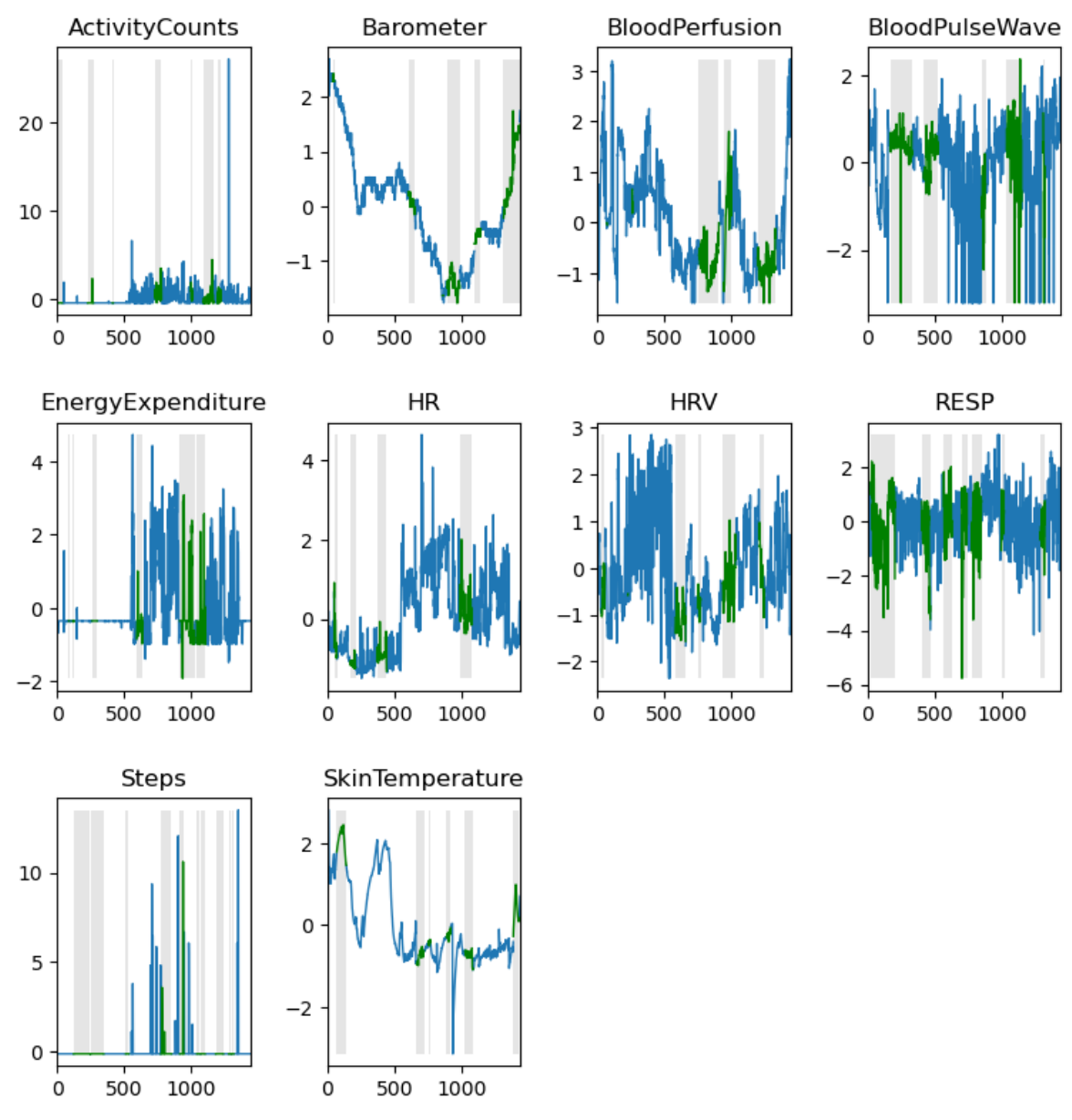}
    % \caption{Example of the 10 physiological and behavioral signals. The X-axis shows the time of the recording and the Y-axis the sensor value. Blue and green colors refer to the true value of the signal and green to the masked missing data. }
    % \label{fig:real_imp_visual}
 \end{minipage}
  \hspace{0.01pc}
 \begin{minipage}[b]{.44\textwidth}
    \centering
    \includegraphics[scale=0.3]{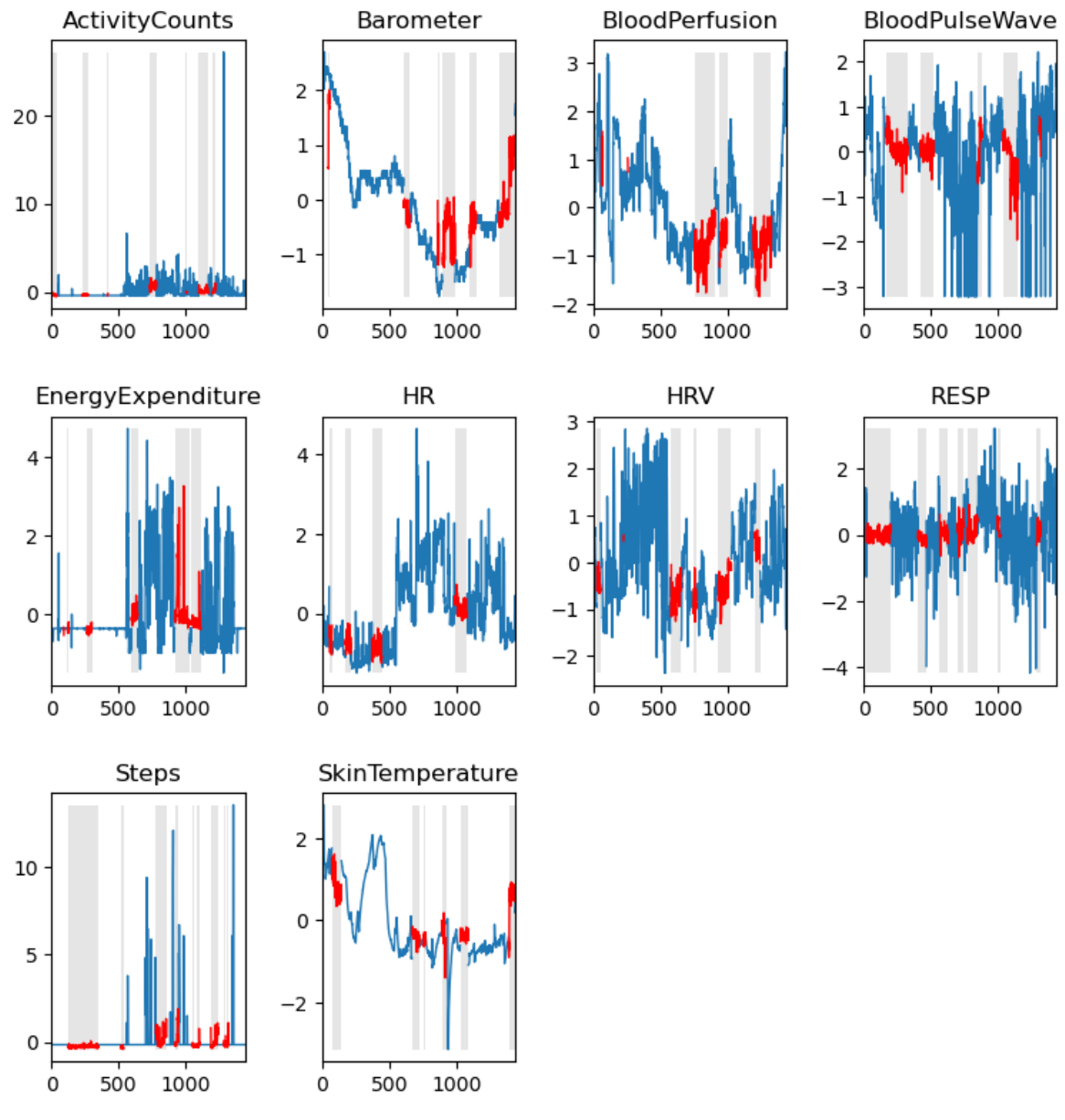}
 \end{minipage}
\begin{minipage}[b]{.44\textwidth}
      \centering
    \includegraphics[scale=0.3]{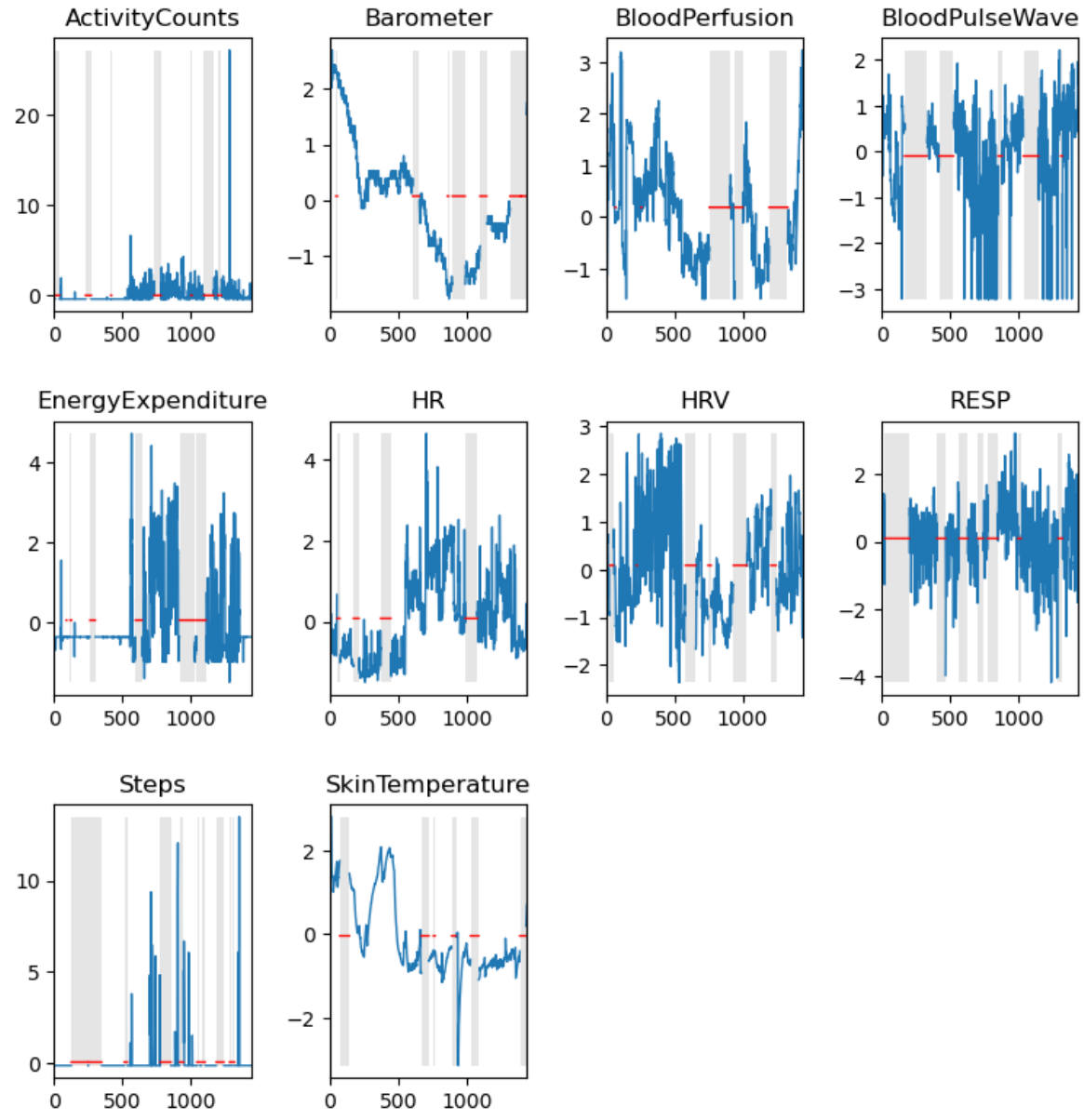}
    % \caption{Examples of the reconstruction of the missing data of 10 parameters using transformer imputation. X-axis shows the time of the recording and the Y-axis the sensor value. Blue color refers to the original trace and red to the imputed values. }
    % \label{fig:mean_imp_visual}
 \end{minipage}
 \hspace{0.01pc}
 \begin{minipage}[b]{.44\textwidth}
    \centering
    \includegraphics[scale=0.3]{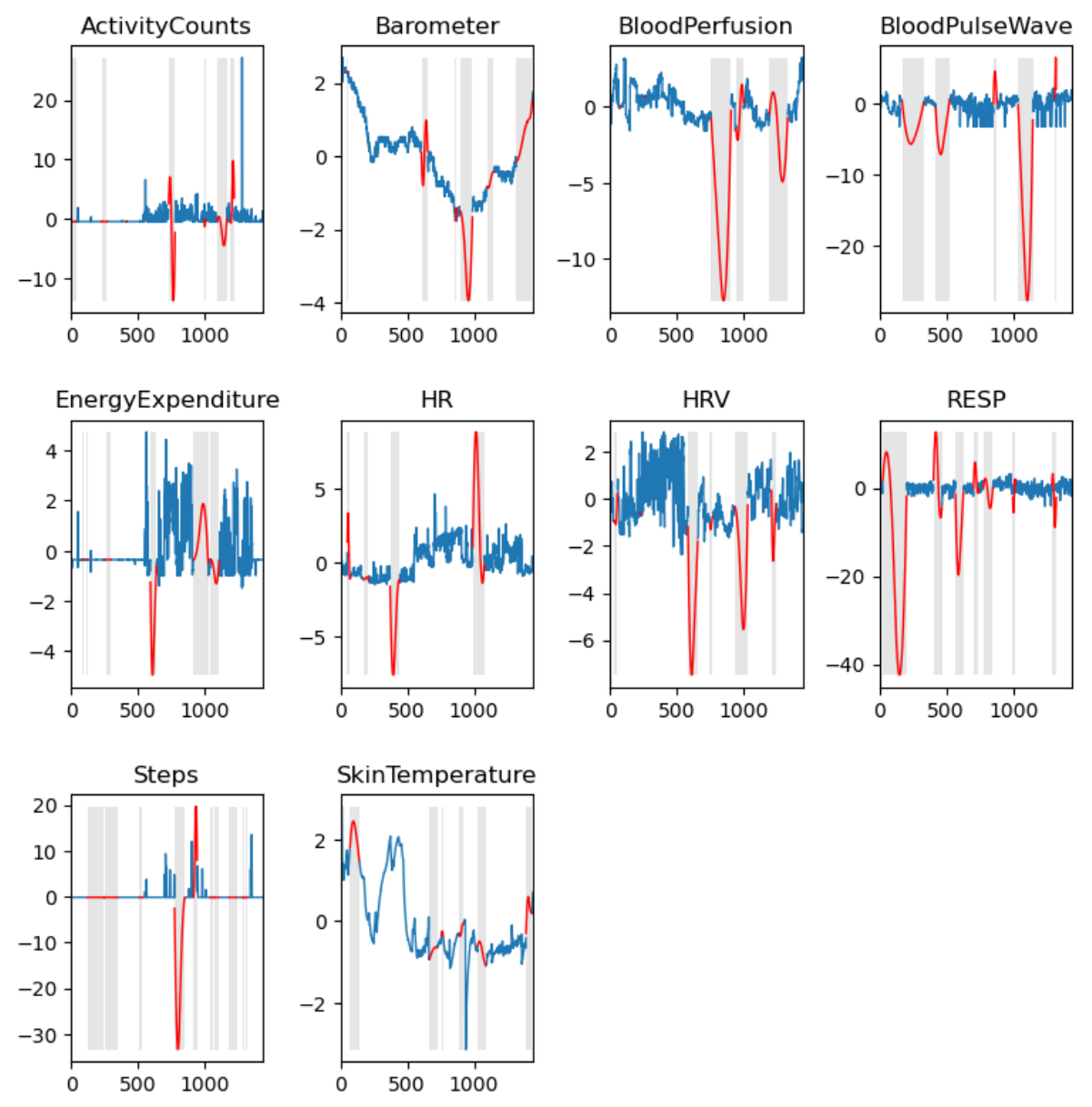}
 \end{minipage}
\begin{minipage}[b]{.44\textwidth}
      \centering
    \includegraphics[scale=0.3]{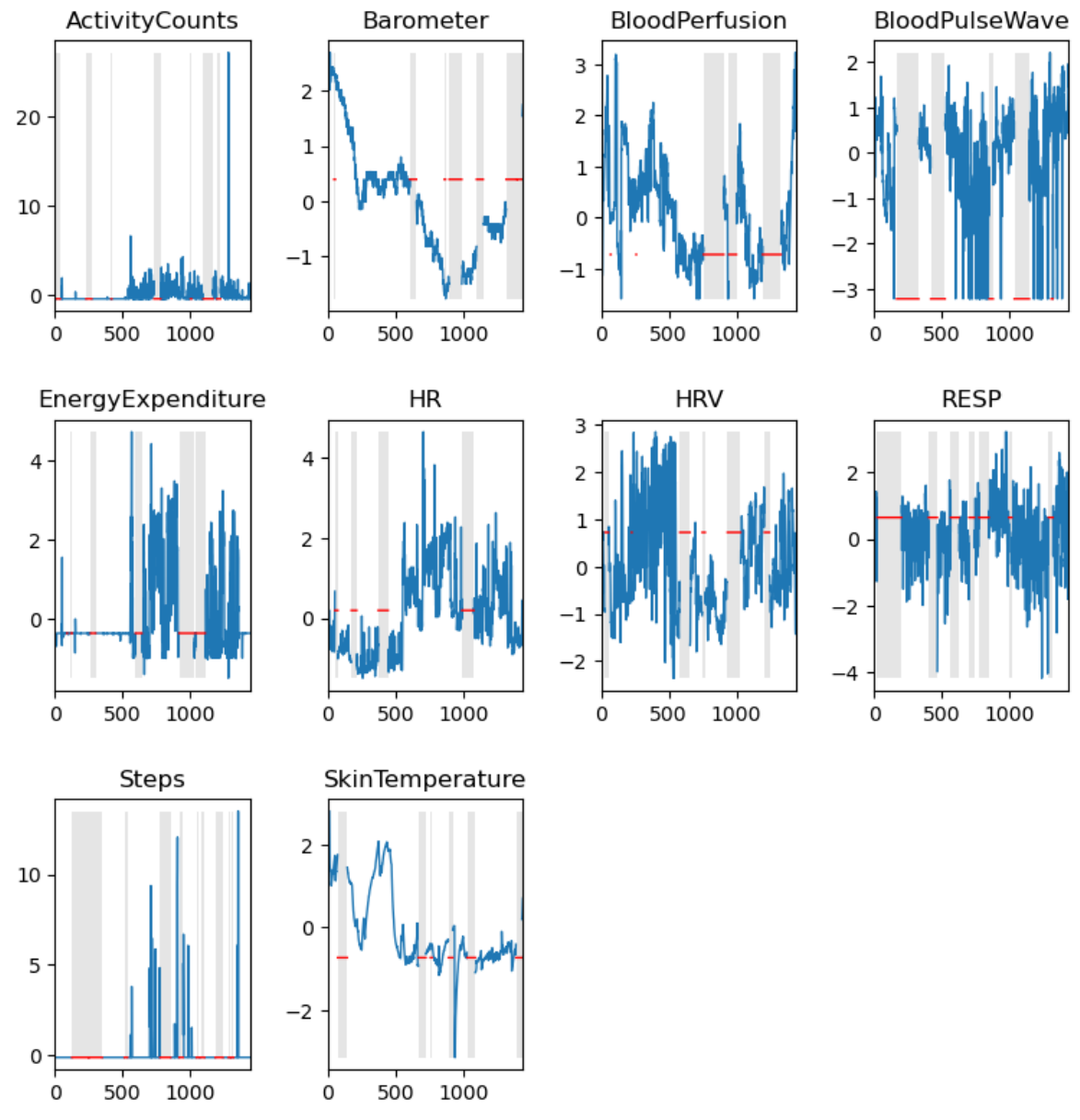}
    % \caption{Examples of the reconstruction of the missing data of 10 parameters using transformer imputation. X-axis shows the time of the recording and the Y-axis the sensor value. Blue color refers to the original trace and red to the imputed values. }
    % \label{fig:mean_imp_visual}
 \end{minipage}
 \hspace{0.02pc}
 \begin{minipage}[b]{.45\textwidth}
    \centering
    \includegraphics[scale=0.3]{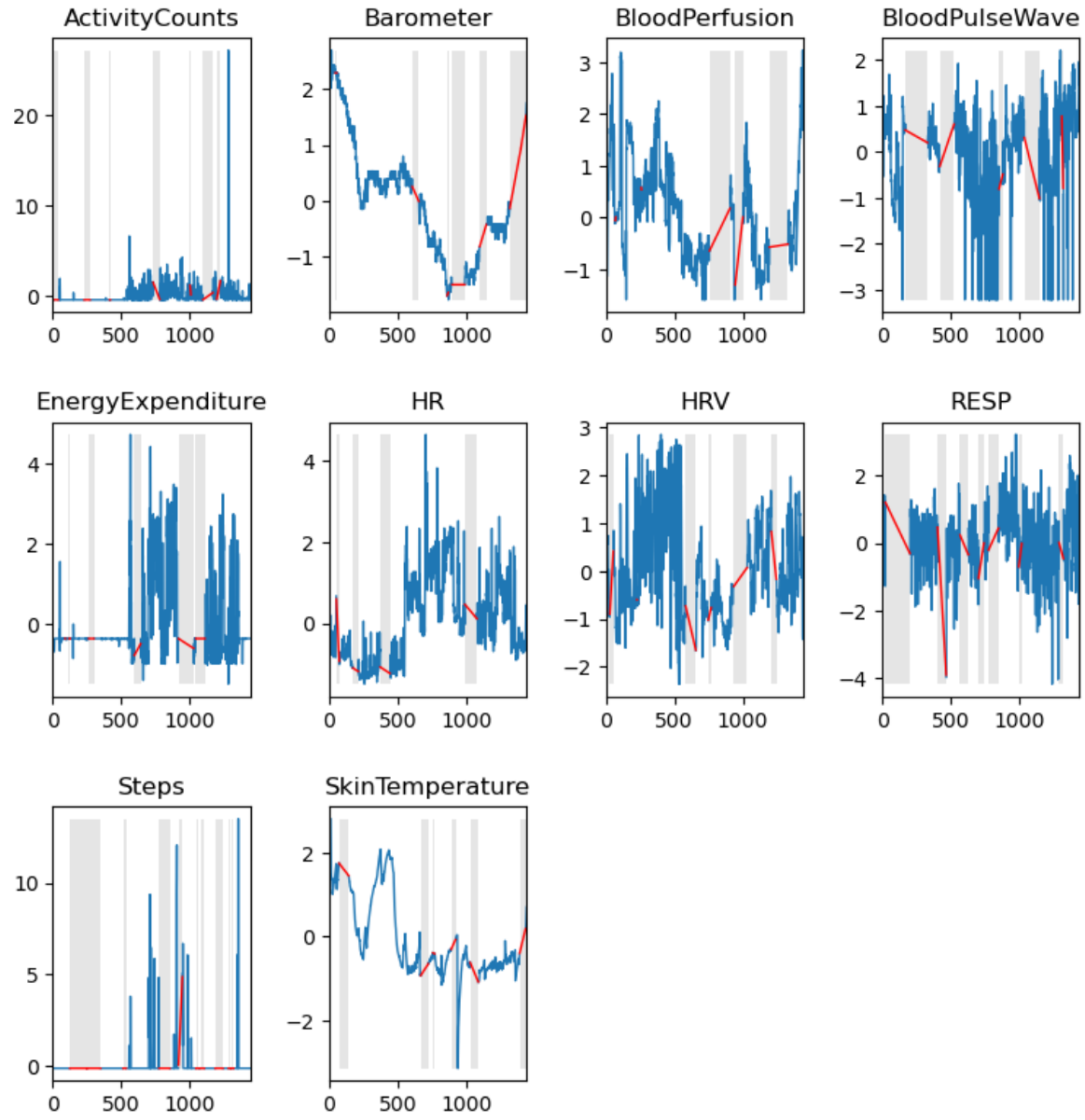}
 \end{minipage}
\caption{Example of the physiological and behavioral signals explored in this work. The X-axis shows the recording time and the Y-axis the sensor value. Blue color refers to the true value of the signal, green to the masked missing data, and red to the imputed data with \textit{transformer} (top-right), \textit{mean} (middle-left), \textit{quadratic} (middle-right), \textit{mode} (bottom-left), and \textit{linear interpolation} (bottom-right) imputation strategies. }
\label{fig:imp_visual}
\end{figure*}

Figure \ref{fig:imp_visual} shows examples of the true (green color) and reconstructed signals (red color) using the transformer, mean, and quadratic imputation strategies on example time windows for each parameter. Overall, the transformer-imputed values seem reasonable in most cases for the majority of physiological parameters. Indeed, it is capable of preserving the patterns of the HR, HRV and EE. We observe that also visually the mode imputation seems to work best for imputing the number of steps.
% The transformer tries to capture the underlying signal patterns and the general trend of the data, which could lead to larger errors for short sequences. 
Overall, the quadratic imputation strategy does not seem appropriate for any of the signals. 

% \section{Ethical statement.}
% You can write a statement about the potential ethical impact of your work, including its broad societal implications, both positive and negative. If included, such statement must be written in an unnumbered section titled \emph{Ethical statement}.

\section{Acknowledgments}

S.G. is supported by the Swiss Data Science Center (SDSC) through the grant C21-18P and a postdoctoral fellowship by the ETH AI Center.

% The acknowledgments section, if included, appears right before the references and is headed ``Acknowledgments". It must not be numbered even if other sections are (use \texttt{\textbackslash section*\{Acknowledgements\}} in \LaTeX{}). This section includes acknowledgments of help from associates and colleagues, credits to sponsoring agencies, financial support, and permission to publish. Please acknowledge other contributors, grant support, and so forth, in this section. Do not put acknowledgments in a footnote on the first page. If your grant agency requires acknowledgment of the grant on page 1, limit the footnote to the required statement, and put the remaining acknowledgments at the back. Please try to limit acknowledgments to no more than three sentences.

\end{document}